\documentclass[10pt,conference]{IEEEtran}
\usepackage{ifpdf}

\ifCLASSINFOpdf
\usepackage[pdftex]{graphicx}
\else
\fi
\ifCLASSOPTIONcompsoc
  \usepackage[caption=false,font=normalsize,labelfont=sf,textfont=sf]{subfig}
\else
  \usepackage[caption=false,font=footnotesize]{subfig}
\usepackage{url}

\usepackage{threeparttable}

\usepackage{pgfplots}
\usepackage{tabularx}
\usepackage{makecell}
\usepackage{multirow}
\usepackage{rotating}
\usepackage{xcolor}

\usepackage[font=small,skip=0.5pt]{caption}

\newcommand{\specialcell}[2][l]{%
  \begin{tabular}[#1]{@{}l@{}}#2\end{tabular}}

\usepackage{listings}
\makeatletter
\def\lst@makecaption{%
  \def\@captype{table}%
  \@makecaption
}
\makeatother

\begin{document}

%
\title{\huge Well Begun is Half Done: An Empirical Study of Exploitability \& Impact of Base-Image Vulnerabilities}





\author{\IEEEauthorblockN{
Mubin Ul Haque\IEEEauthorrefmark{2} and
M. Ali Babar \IEEEauthorrefmark{3}}
\IEEEauthorblockA{Centre for Research on Engineering Software Technologies (CREST)}
\IEEEauthorblockA{School of Computer Science, and Engineering, The University of Adelaide, Adelaide, Australia}
\IEEEauthorblockA{Cyber Security Cooperative Research Centre, Australia}
\IEEEauthorblockA{ \IEEEauthorrefmark{2}mubinul.haque@adelaide.edu.au, \IEEEauthorrefmark{3}ali.babar@adelaide.edu.au}
}


%


\maketitle
\thispagestyle{plain}
\pagestyle{plain}

\begin{abstract}
Container technology, (e.g., Docker) is being widely adopted for deploying software infrastructures or applications in the form of container images. 
Security vulnerabilities in the container images are a primary concern for developing containerized software. 
Exploitation of the vulnerabilities could result in disastrous impact, such as loss of confidentiality, integrity, and availability of  containerized software. 
Understanding the exploitability and impact characteristics of vulnerabilities can help in securing the configuration of containerized software.  
However, there is a lack of research aimed at empirically identifying and understanding the exploitability and impact of vulnerabilities in container images. 
We carried out an empirical study to investigate the exploitability and impact of security vulnerabilities in base-images and their prevalence in open-source containerized software. 
We considered base-images since container images are built from base-images that provide all the core functionalities to build and operate containerized software. 
Besides, security vulnerabilities in a base-image can propagate to derived container images, which can host different applications. That means a single exploitable vulnerability in base-images can result in security attacks in several containerized software.
We discovered and characterized the exploitability and impact of security vulnerabilities in 261 base-images, which are the origin of 4,681 actively maintained official container images in the largest container registry, i.e., Docker Hub. 
To characterize the prevalence of vulnerable base-images in real-world projects, we analysed 64,579 containerized software from GitHub.
Our analysis of a set of $1,983$ unique base-image security vulnerabilities revealed 13 novel findings.
These findings are expected to help developers to understand the potential security problems related to base-images and encourage them to investigate base-images from security perspective before developing their applications. 
For researchers, this study highlights the need of developing tools for mitigating the exploitability of vulnerable base-images.

\end{abstract}

\textbf{Keywords-} Base-image, Container, Docker, Vulnerabilities

%
\IEEEpeerreviewmaketitle

\section{Introduction}
\vspace{- 5 pt}
Containers become increasingly popular as a virtualization technology to build softwarized infrastructures. Portworx survey in 2019 suggested 87\% enterprises were using containers, compared to 67\% in 2017 \cite{portworx}. Moreover, the container market was predicted to expand from 1.2 billion in 2018 to 4.9 billion (USD) by 2023 \cite{market}. A container can be defined as a stand-alone executable unit of software that includes an entire environment \cite{dockerdocks}, which promotes modularity and reproducibility, two of the reasons for containers being popular \cite{stackrox}. 

Docker is one of the most widely used container-based technology. Docker had been ranked first in  \textit{``Most Wanted Platform''}, second in \textit{``Most Loved Platform''} and third in \textit{``Platform In Use''} in the consecutive annual survey of Stack Overflow with the participation of almost 90,000 and 65,000 developers around the world in 2019 and 2020, respectively \cite{stackOverflow20}. These survey results are a clear indication of developers' massive interest in Docker. Docker images provide a convenient way to store and deliver containerized software. 

While adopting Docker technology, security vulnerabilities in container images are critical consideration, since the exploitation of vulnerabilities can negatively affect the confidentiality, integrity, and availability of containerized software \cite{shu2017study,mohallel2016experimenting,martin2018docker}. 
Moreover, \textit{StackRox}, \textit{Amazon}, and \textit{Redhat} reported several survey results where developers ranked vulnerability management as the top-concern in establishing a reliable container environment \cite{stackrox},  \cite{amazon}, \cite{redhat}. Vulnerability management in containerized software includes various steps, such as detection, assessment, prioritization, and remediation of vulnerabilities \cite{martin2018docker}. 
Vulnerability assessment and prioritization is one of the key steps for vulnerability management, which includes the determination of exploitability characteristics, such as attackers' skill, location (e.g., external network or local network), requirement (e.g., privileges or interaction from the user of containerized software) to exploit the vulnerability and impact characteristics, such as attackers' target to breach data confidentiality, data integrity, and data availability \cite{lin2018measurement, YIN2020106529}. 

This assessment and prioritization is indispensable for practical realization of security configuration for containerized software. For example, a containerized software contains a `Out of Bound Write' vulnerability (e.g., CVE-2019-19319) where the attacker required admin privileges to  exploit the vulnerability. However, the exploitation of this vulnerability can be minimized if the system administrator restricts the root privileged option for the users of such software. Therefore, vulnerability assessment and prioritization can help hardening containerized software configuration based on various exploitability and impact characteristics.

Previous studies have identified the distribution and types of vulnerabilities \cite{shu2017study,zerouali2019relation,wist2020vulnerability,ibrahim2020too}, the existence of vulnerabilities in different image repositories \cite{socchi2019deep,zerouali2021impact,zerouali2021multi}, vulnerable packages and rule violations \cite{tak2018security} in Docker Hub (DH) images. While these previous studies focused on detection step of vulnerability management of container images, there is a research gap to investigate the assessment and prioritization of container image vulnerabilities. Our research aims to fill this gap by conducting an empirical study on the assessment and prioritization of container image vulnerabilities. 

The goal of our research is to provide an evidence-based knowledge and understanding of the exploitability, impact, and prevalence of container image vulnerabilities. We conducted the study on the container base-image vulnerabilities, since all the containerized software are derived from base-images. A base-image can be defined as a referential image, used to build custom images containing software \cite{dockerdocks}. In fact, base-images provide the main support in terms of functionality to build and develop containerized software.

We analyzed 261 base-images, upon which 4,681 actively maintained container images are being hosted in DH using ANCHORE \cite{anchoreTool} to identify their vulnerabilities. Then we crawled exploitability and impact metrics for each of the identified vulnerability from National Vulnerability Database \cite{nvd}. We also studied the Proof of Concept (PoC) exploitation of base-image vulnerabilities, which had been collected from Exploit Database (EDB) \cite{exploitdb}, VulHub \cite{vulhub}, and Metasploit \cite{metasp}. Besides, 64,579 GitHub (GH) projects were investigated to understand the practice of using base-images to build and develop containerized software. Our empirical endeavour in this study has enabled us to determine 13 novel findings. Some of our findings include: 
\begin{itemize}
    \item highly exploitable vulnerabilities existed more in minimal base-image repositories, 
    \item exploitation of large operating system base-images have more severe impact in terms of confidentiality, integrity, and availability,
    \item vulnerabilities with PoC exploit are prevalent in actively updated and official DH base-images, 
    \item GH projects maintained by organizations or having large teams are more likely to use secure base-images.
\end{itemize}
Our knowledge-base for base-image vulnerabilities can help developers to learn about how to identify and avoid vulnerable base-images and researchers to develop suitable techniques and tools to discover and minimize the exploitation of base-image vulnerabilities. In particular, knowing the characteristics of exploitation and impact of base-image vulnerabilities is expected to guide (1) container developers, to understand the risks of using base-images without first identifying and addressing their exploitability during the build, prepare, and deployment phase, (2) Docker developers, in focusing their verification and validation efforts towards building secure base-images, (3) researchers, to design/improve the process for selection of base-images to ensure secure compliance of containerized infrastructures.

The remainder of our paper proceeds as follows. Section \ref{section: background} and \ref{sec: threatModel} describe the research background, problem, and threat model. Sections \ref{method} and \ref{sec: results} discuss the research approaches and the process of the empirical investigation. We report the usefulness, implications, limitation, and related work in Section \ref{section: impact}, \ref{sec: discussion}, \ref{sec: threats}, and \ref{sec: related}, respectively. Finally, our paper concludes in Section \ref{sec: conclussion} with some future research directions.

\section{Research Background and Research Problem}
\label{section: background}
\subsection{Docker Images and Base-Images} 
An application encapsulated by Docker is distributed in the form of an image \cite{dockerdocks}. Images are file systems, which are composed of \emph{layers}. Each \emph{layer} of the image is a representation (or an outcome) of an instruction, when an instance of an image is executed. These instructions are often described in a Dockerfile, which is a text-based file that specifies all the needed information about the environment and program execution. 
There exists a parent-child relationship in an image \cite{dockerdocks}. A base-image is the root of this inheritance tree, providing bare-bone functionality of a specific platform to run images. To build container images, we need a starting point, which is indeed necessary as Docker uses a Union File System, and the whole file system is mounted read-write \cite{dockerdocks}. All the changes occur to the top-most layer, essentially built upon base-images. 

\vspace{- 5 pt}
\subsection{Research Problem and Research Questions}
\label{section: reseachProblem}
\begin{figure}[]
    \centering
    \includegraphics[width = 3.45 in,  ]{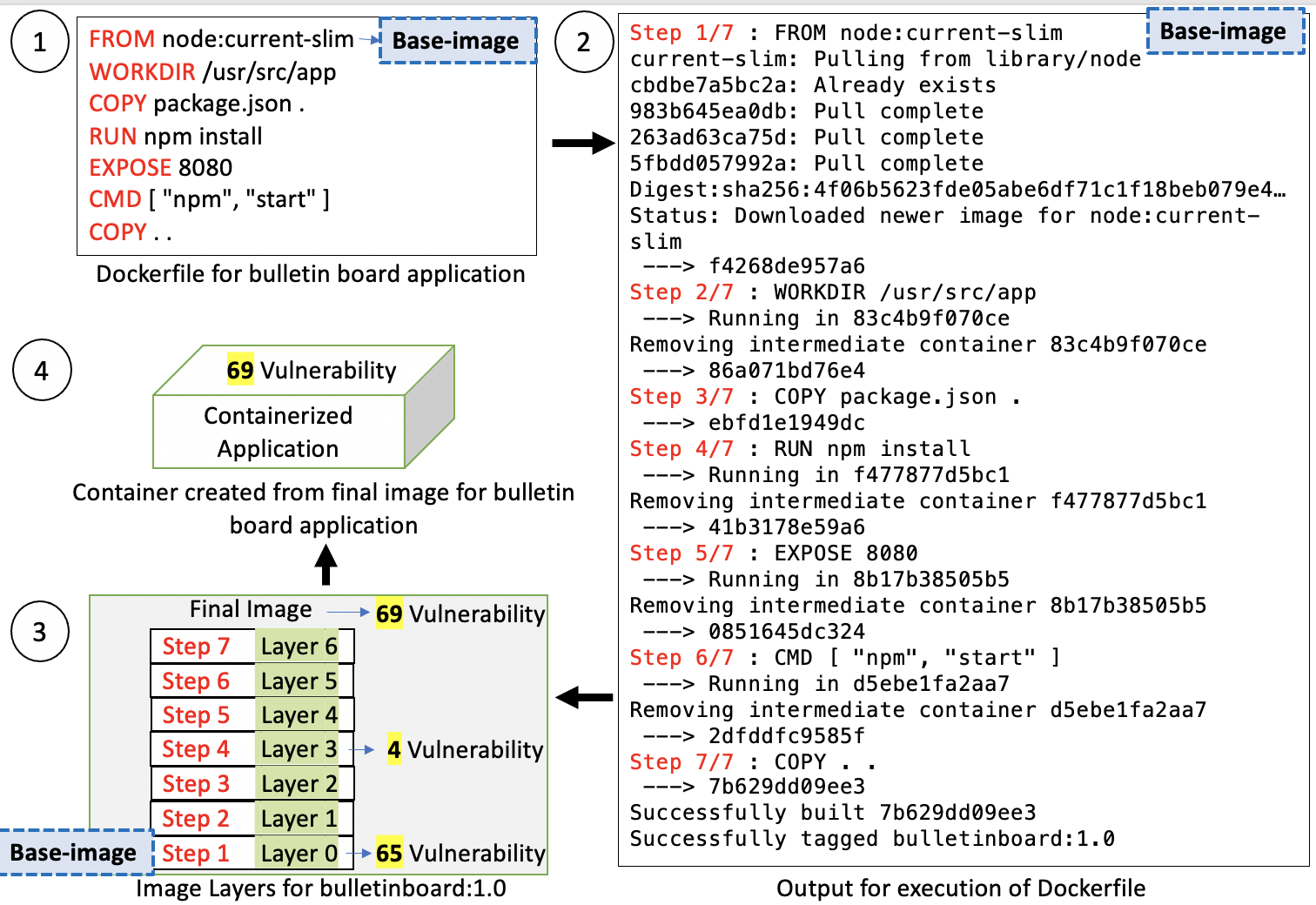}
    \caption{Vulnerability propagation from base-image to application.}
    \label{fig: vulnProp}
\end{figure}

As previously stated, vulnerabilities can be propagated from a base-image to a final image used for a containerized software. To illustrate an example, we considered the bulletin-board software, provided in Docker Documentation \cite{nodebulletinboard}. To build  the software or application image, \textit{node:current-slim} was used as a base-image, which is shown as \textcircled{1} in Fig.  \ref{fig: vulnProp}. We built the application image using \texttt{docker build} command, whose execution is shown as \textcircled{2} in Fig. \ref{fig: vulnProp}. We tested the application image based on \textit{node:current-slim} using ANCHORE \cite{anchore}. The tool reported 69 vulnerabilities for the application image. We investigated the image building process by leveraging the Dockerfile  of \textit{node:current-slim} and found that it built on top of the image \textit{debian:stretch-slim} \cite{debstretch}. We further investigated the Dockerfile of \textit{debian:stretch-slim} and found that it had been built on top of \textit{scratch} image \cite{scratchDockerfile}. Then we identified the vulnerability using ANCHORE for each of \textit{scratch}, \textit{debian:stretch-slim} and \textit{node:current-slim}. By investigating the vulnerable packages among the images, we found that 65 vulnerable packages of \textit{node:current-slim} and \textit{debian:stretch-slim} were identical whereas \textit{scratch} had been found with no vulnerabilities since \textit{scratch} is an empty image \cite{scratchDockerfile} and only used as a foundation to build other images. Notably, all of these 65 vulnerabilities were propagated to the application image of bulletin-board, which is exhibited as \textcircled{3} in Fig. \ref{fig: vulnProp}. 

The remaining four vulnerabilities were obtained during the installation of the packages. Finally, 69 vulnerabilities existed in the containerized software created from the application image using the \texttt{docker run} command, depicted as \textcircled{4} in Fig. \ref{fig: vulnProp} where 94.2\% of the vulnerabilities were inherited from the base-image \textit{node:current-slim}. 
The hierarchical relationship between a base-image layer and the subsequent layers makes the propagation of vulnerabilities inevitable \cite{stackrox, redhat, socchi2019deep}. Therefore, the vulnerabilities in a base-image are likely to show up, exploit and leave impact on all of the applications (built upon that base-image).

Besides, prior research \cite{leetal} discussed qualitative severity measurement, such as `high/medium/low' \cite{cvss} may often fail to indicate the exploitability and impact of the vulnerability, which is essential for assessing the security risk of the software. For example, CVE-2019-14899 (Improper access control) is a `high' severity vulnerability; however, it requires admin access, some interaction from a user to initiate the attack for exploitation, and the network to be local, indicating this vulnerability can be exploited by few people. On the other hand, vulnerabilities with medium or low severity can be exploited at first to gain unauthorized access and then used to perform malicious activity in the victim software \cite{YIN2020106529, bilge2012before, joh2011defining}.
Therefore it is important to know and understand the exploitability and impact of the known vulnerabilities in base-images so that containers developers can select secure base-images before developing applications in a containerized context. Our ongoing research, including this study, in this area is aimed at revealing, characterizing, and providing a suitable knowledge-base for the exploitability and impact of base-image vulnerabilities since they are the potential source of a majority of the vulnerabilities in containerized software. 

Our study aims to answer following Research Questions (RQs): 
\subsubsection{\textbf{RQ1}} \textit{What is the exploitability characteristics of base-image vulnerabilities?}
Each container running on a machine shares the same operating system kernel \cite{sultan2019container} and containers are originated from the base-images. Given base-image security vulnerabilities can be propagated to the containers, exploitable vulnerabilities could breach the host operating system, and thus all containers in a machine would be compromised due to exploitation of base-image vulnerabilities. Through the \textbf{RQ1}, we intended to investigate the current exploitability landscape, e.g., location, requirement, and capability of the attacker to exploit the known vulnerabilities in actively used base-images of DH. An empirically derived answer to this question is important for increasing the knowledge about the security vulnerabilities exploitation that may exist in base-images; and the skills or efforts required to exploit such vulnerabilities. The knowledge from the answer to \textbf{RQ1} is expected to help practitioners to prioritize their efforts for securing configuration in containerized software.
\subsubsection{\textbf{RQ2}} 
\textit{What is the impact of the identified vulnerabilities?} Understanding the consequences of the vulnerabilities is important for developers to protect their assets from the outcomes of a successful exploitation of vulnerabilities. An answer to the \textbf{RQ2} will help practitioners to learn about the potential impact, e.g., confidentiality, integrity, and availability of different base-images vulnerabilities so that they are in a position to select suitable base-images for their development. For instance, banking transaction system developers can prioritise confidentiality and integrity of the data over the availability of the application interface \cite{leetal} and thus, can select base-images with lower or no impact in terms of confidentiality and integrity. Researchers can be motivated to investigate the approaches for minimizing the impact of the base-image vulnerabilities exploitation. 
\subsubsection{\textbf{RQ3}} 
\textit{To what extent do the identified vulnerable base-images are deployed in open-source projects?} We aim to explore how many of the vulnerable base-images are being deployed in open-source projects to host containerized software. The knowledge gained from such identification will help to understand and characterize the current practices of using base-images in containerized software development. An answer to the \textbf{RQ3} may encourage practitioners to use the secure base-images in their containerized software and motivate researchers to focus more on developing suitable techniques and tools for ensuring the security of base-images.
   
\section{Threat Model}
\label{sec: threatModel}
The purpose of this section is to discuss the threat model by conceptualizing the container operation, corresponding attack goal, and capabilities of attackers. The goal of the attacker can be disclosure of sensitive information, such as passwords, application credentials, or secrets, data tampering or resource exhaustion. In our threat model, we assume the resource providers and image owners as the trusted entities, i.e., attackers can not conspire with them. In other words, we are not considering any local adversary who has a full access to a container because the image owner or resource provider is capable of achieving any adversary effects (i.e., breaking the container isolation or resource exhaustion) without any vulnerabilities in the base-image. Therefore, we assume the attackers could only penetrate from external or internal network. 


\textbf{Attack purposes.} Attackers’ targets might be the containers or the other container-based applications hosted on them to steal sensitive data or classified information. Another common motivation is to gain access to personally identifiable information,  such as healthcare information, or bio-metrics which can be leveraged to commit insurance or credit card fraud. Moreover, attackers can overload data centers with a Distributed Denial of Service (DDoS) attack to cause downtime, harm business process, and incur financial loss. 

\textbf{Attackers' capabilities.} We analysed and explained the following aspects of attackers' capabilities. 

\textbf{Internal attacker.} A container can be targeted from a remote server in the internal network. We assume that attacker resides in the same network of the target container, but does not have any direct access to  target container \cite{modi2017virtualization, arjunan2017enhanced}. 
\textbf{External attacker.} An external attacker  requires remote network access to the target container \cite{modi2017virtualization, arjunan2017enhanced, kong2020automated}. Here, we assume an attacker does not have any privileges to access the container. 

\section{Research Approaches and Process}
\label{method}
We describe the research approaches used and the process followed by this study in this section.

\subsubsection{\textbf{Building Base-image specific data set}} 
\begin{figure}[]
    \centering
    \includegraphics[width=\linewidth]{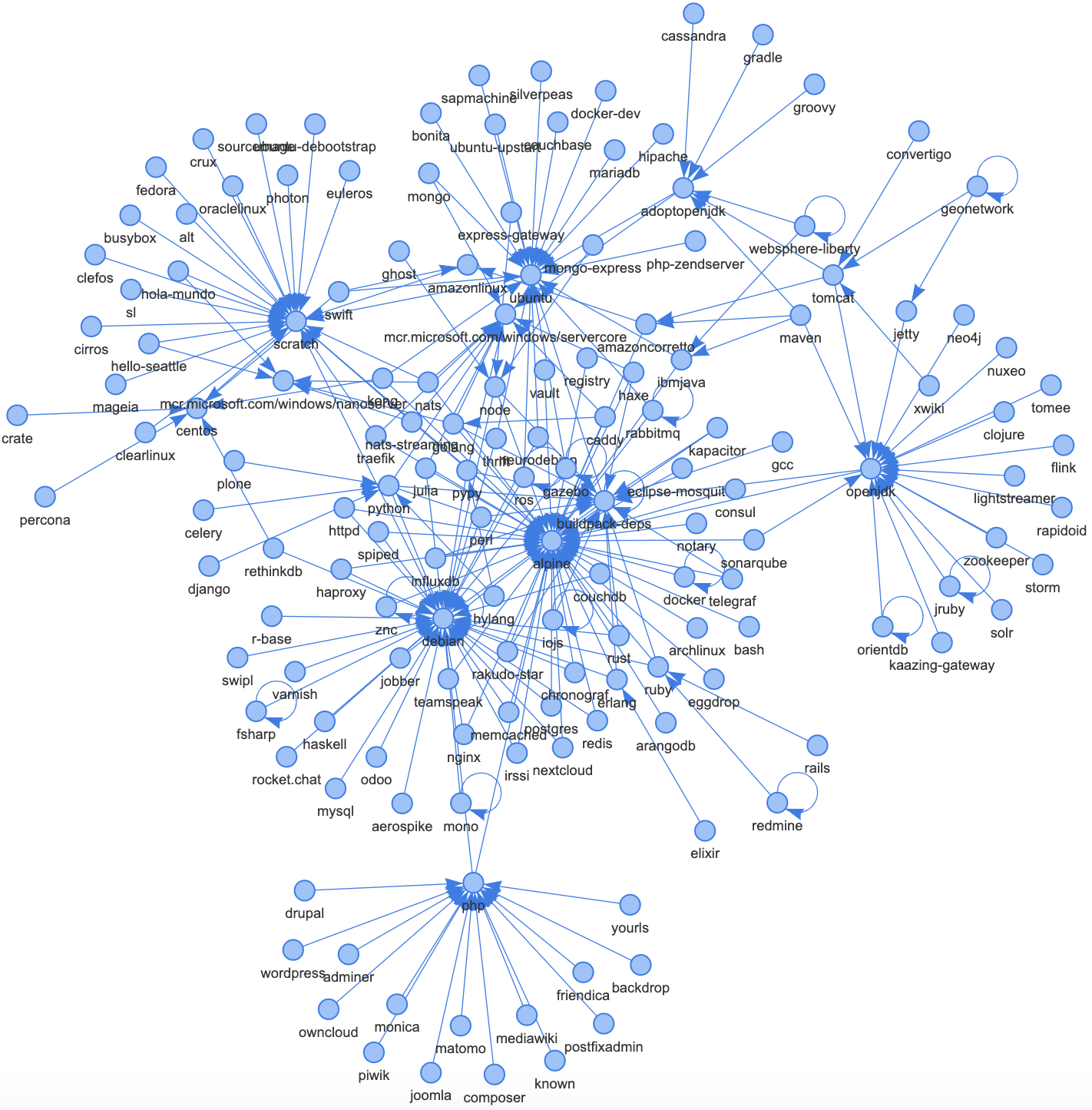}
    \caption{Dependency among official repositories in DH.}
    \label{fig: depOfficial}
\end{figure}

As a first step towards answering our research questions, we built a data set of base-images which was determined and extracted from DH.
Unlike previous studies \cite{wist2020vulnerability, socchi2019deep}, we decided about base-images on its deployment in DH since tag-based extraction does not provide transitive dependencies among images. For example, image \textit{php:7.3.27-fpm-alpine} has the tag \textit{7.3.27-fpm-alpine}, which does not indicate the image upon which it is being instantiated nor the image it instantiates. In this regard, we considered all the official repositories, and mined 4,681 actively maintained images (using DH API \cite{api}) from 163 repositories as of May 2021. We considered official type image repositories since they are maintained by the Docker team and assumed to be more secured than verified, community or certified types of repositories \cite{shu2017study, zerouali2019relation}. 
An image is considered as a base-image if it directly originates from \textit{scratch}, which is an empty image, and thus does not create any extra layer in the filesystem \cite{scratchDockerfile}. For example, image \textit{php:7.3.27-fpm-alpine} is based on \textit{alpine:3-13}, and \textit{alpine:3-13} is extended upon \textit{scratch}. We considered \textit{alpine:3-13} as the base-image since \textit{alpine:3-13} was built upon \textit{scratch}. Fig. \ref{fig: depOfficial} shows the transitive dependency among official repositories of DH. As shown in Fig. \ref{fig: depOfficial}, base-image repositories are either originated from \textit{scratch}, or one of the images in base-image repositories. Finally our data set was composed of 261 base-images from 8 different repositories (\textit{alpine}, \textit{bash}, \textit{busybox}, \textit{centos}, \textit{debian}, \textit{ibmjava}, \textit{oraclelinux}, and \textit{ubuntu}) where a strong representation of Linux distributions, such as \textit{ubuntu}, \textit{debian}, \textit{alpine}, and \textit{centos} were observed. Table \ref{tab: topBaseImageDockerHub} and \ref{tab: popularity} represent top base-images along with popularity (star and downloads) and maintainability (maintainer and contributor) in DH. 

\begin{table}[]
\centering
\caption{Top-5 base-images to host official images in DH}
\label{tab: topBaseImageDockerHub}
\scalebox{0.8}{
\begin{tabular}{|c|c|c|c|c|}
\hline
Base-Image            & \# dependent images \\ \hline
debian:buster-slim    & 386                \\ \hline
alpine:3.13           & 313                \\ \hline
alpine:3.12           & 234                \\ \hline
ubuntu:16.04 & 132                \\ \hline
debian:stretch-slim   & 101                \\  \hline
\end{tabular}
}
\end{table}

\begin{table}[]
\centering
\caption{Popularity and maintainability of base-image repositories}
\label{tab: popularity}
\scalebox{0.8}{
\begin{tabular}{|c|c|c|c|c|c|c|c|c|}
\hline
Repository  & alpine                                              & bash                                                & \begin{tabular}[c]{@{}c@{}}busy-\\ box\end{tabular} & centos                                              & debian                                              & \begin{tabular}[c]{@{}c@{}}ibm-\\ java\end{tabular} & \begin{tabular}[c]{@{}c@{}}oracle-\\ linux\end{tabular} & ubuntu                                              \\ \hline
Star        & 7634                                                & 385                                                 & 2285                                                & 6628                                                & 3907                                                & 90                                                  & 774                                                     & 12481                                               \\ \hline
Downloads   & \begin{tabular}[c]{@{}c@{}}4.93E\\ +09\end{tabular} & \begin{tabular}[c]{@{}c@{}}2.17E\\ +08\end{tabular} & \begin{tabular}[c]{@{}c@{}}4.16E\\ +09\end{tabular} & \begin{tabular}[c]{@{}c@{}}9.15E\\ +08\end{tabular} & \begin{tabular}[c]{@{}c@{}}6.94E\\ +08\end{tabular} & \begin{tabular}[c]{@{}c@{}}1.14E\\ +07\end{tabular} & \begin{tabular}[c]{@{}c@{}}2.39E\\ +07\end{tabular}     & \begin{tabular}[c]{@{}c@{}}3.86E\\ +09\end{tabular} \\ \hline
Maintainer  & 1                                                   & 1                                                   & 3                                                   & 2                                                   & 1                                                   & 1                                                   & 1                                                       & 2                                                   \\ \hline
Contributor & 4                                                   & 3                                                   & 3                                                   & 4                                                   & 1                                                   & 2                                                   & 4                                                       & 7                                                   \\ \hline
\end{tabular}
}
\end{table}
\subsubsection{\textbf{Vulnerability Scanning}}
\label{sectest}
Once the base-image names and their tags had been identified, we performed scanning of the base-images using ANCHORE \cite{anchoreTool}, an open-source tool for vulnerability identification and also used by previous researches \cite{wist2020vulnerability, liu2020understanding, javed2021understanding}. ANCHORE was chosen for several reasons. Firstly, ANCHORE uses a number of sources for vulnerability databases which represent rich vulnerability feed for ANCHORE. For instance, ANCHORE uses security advisories from specific Linux distribution vendors against distribution specific packages. Alpine, CentOS, Debian, Oracle Linux, Red Hat Enterprise Linux, Ubuntu, Amazon Linux, and Google Distroless are included in Linux distribution vendors \cite{anchorefeed}. ANCHORE also consults National Vulnerability Database (NVD) \cite{nvd} and GitHub Advisory Database \cite{githubadv} to provide rich vulnerability information. In contrast, other open-source container security tools, such as Trivy \cite{trivy}, Dagda \cite{dagda}, and Clair \cite{clair} use a subset of the above-mentioned sources. For example, Trivy, Dagda, and Clair do not mention about using GitHub advisory databases and Google Distroless distribution. Secondly, Berkovich \textit{et al.} showed that ANCHORE reported lower false negatives/positives and higher true positives than other security tools \cite{berkovich2020ubcis}. Thus, ANCHORE provides better precision, recall, and F-measure than Trivy and Clair \cite{berkovich2020ubcis}. Third, ANCHORE downloads data from sources in every 6 hours \cite{anchoredownload} which indicates regular synchronization of vulnerability feed. Authors of Dagda \cite{dagda}, Trivy \cite{trivy},  and Clair \cite{clair} do not mention about synchronization of vulnerability feed. Regular synchronization of feed from sources is vital for updating vulnerability information \cite{berkovich2020ubcis}.



ANCHORE identifies vulnerabilities by matching the components (installed packages, executable, libraries, and operating system specific files in images) with security feeds for applicable vulnerabilities. ANCHORE provides the following types of vulnerability information.  
\begin{enumerate}
    \item Vulnerability ID: Common Vulnerabilities and Exposure (CVE) identifier to detect vulnerability. 
    \item Package: Names and exact version of the associated packages with each vulnerability.
    \item Fix: Name and version of the package which fix the identified vulnerability. If there is no such fix, ANCHORE provides `none'. We leveraged this metric to identify the availability of the patch or fix. 
\end{enumerate}

We obtained 2,269 unique security vulnerabilities for our base-image data set. 


\subsubsection{\textbf{Exploitability and Impact metrics}} 
\label{sec: explotMetrics}  To answer RQ1 and RQ2, exploitability and impact metrics for each of the identified vulnerabilities were collected from National Vulnerability Database (NVD) \cite{nvd}. NVD is considered standard practices and verified by security experts \cite{leetal}. Besides, we used the most recent version of Common Vulnerability Scoring System (version 3 \cite{cvss}) as of May 2021 for extracting exploitability and impact metrics. NVD defines exploitability as the reflection of the properties of the vulnerability that led to a successful attack. Exploitability metrics is further composed of \textit{attack complexity, privileges required, user interaction} and \textit{attack vector}. 

\begin{itemize}
    \item \textit{Attack Complexity} (AC) describes the attackers' conditions that must exist in order to exploit the vulnerability. Two possible values for AC are `Low' (L), representing specialized circumstances are not required by attacker and `High' (H), indicating  a successful attack requires the attacker to invest in some measurable amount of effort in preparation or execution against the vulnerability.
    \item \textit{Privileges Required} (PR) describes the level of privileges an attacker requires to exploit the vulnerability. Three possible values for PR are `None' (N), `Low' (L), and `High' (H). `None' (N) denotes attackers do not require access to settings or files to carry out an attack and `Low' (L) indicating a successful attack requires the attacker to posses basic user capabilities that could normally affect only settings and files. `High' (H) denotes attackers require privileges that provide significant (e.g., administrative) control over the vulnerable container.
    \item \textit{User Interaction} (UI) describes the requirement for a human user (or user-initiated process), other than the attacker, to participate in the successful exploitation of the vulnerability. Two possible values for UI are `None' (N), representing the vulnerability can be exploited without any human interaction, `Required' (R), indicating an exploitation requires some form of user interaction.
    \item \textit{Attack Vector} (AV) describes the path or means by which an attacker can gain unauthorized access. Four possible values for AV are `Network' (Net), representing the vulnerability can be exploited through remote network; `Adjacent' (Ad), indicating an attack must be launched from the same shared physical or logical network of vulnerable container, `Local' (L) denoting  attacker’s exploitation path is via read, write or execute capabilities; `Physical' (P) means the attack requires the attacker to physically access or manipulate the vulnerability.
\end{itemize}
We used the prior research \cite{patil2019designing} definition for High likely Exploitation (HE) metrics, composed of `Low' \textit{AC}-`None' \textit{UI}-`None' \textit{PR}. Moreover, impact metrics  captured the effects of a successfully exploited vulnerability on the vulnerable component associated with attack \cite{nvd} and defined in terms of \textit{confidentiality, integrity}, and \textit{availability} or referred to as \textit{CIA} triad.
\begin{itemize}
    \item \textit{Confidentiality} (C) describes the attackers' ability to access information in the exploited container. 
    \item \textit{Integrity} (I) refers to the attackers' ability to modify information in the exploited container.  
    \item \textit{Availability} (A) indicates the impact on the accessibility of the exploited container.
\end{itemize}
All of \textit{CIA} metrics have three possible values: `None' (N) denoting no loss or harm, `Low' (Low) meaning the attacker does not have control over the obtained information or resource, and the amount of loss is limited, whereas `High' (H) referring to the complete loss of information, or resources, and cause detrimental impact on affected container. In addition, vulnerabilities consisting of `High' values in each \textit{CIA} metrics comprise the High Impact (HI) metrics, as defined by \cite{patil2019designing}.

It is to be noted that NVD did not have exploitability and impact metrics for 286 vulnerabilities as of May, 2021. Therefore, our study could not study the exploitability and impact of 286 base-image vulnerabilities.  


\subsubsection{\textbf{Filtering of Vulnerabilities}} 
The purpose of filtering the identified vulnerabilities is to provide the relevancy and context, upon which circumstances, the vulnerabilities can be exploited. In this regard, we filtered the vulnerabilities based on the threat model provided in Section \ref{sec: threatModel}. We considered the capability of an attacker as external if \textit{AV} (crawled from NVD, see Section \ref{sec: explotMetrics}) is represented by the value `Network' \cite{patil2019designing}. Similarly, the capability of an internal attacker was characterized with the value of `Local' or `Adjacent' in \textit{AV} \cite{patil2019designing}. However, we did not consider the vulnerabilities under the \textit{AV} with `Physical' value in our discussion of RQs since it required physical access to the container, or running servers. In other words, vulnerability exploitation in this \textit{AV} requires hardware manipulation, i.e., access to the hardware or infrastructures. Since our threat model considered resource or infrastructure providers as a trusted entity, we left the vulnerabilities with `physical' attack vector. Table \ref{tab: exploitabilityDockerHub} shows the distribution of identified vulnerabilities for all possible 48 combinations for exploitability metrics. It is evident that maximum number of vulnerabilities (26.6\%) lies in ``\textit{AV} Network, \textit{AC:} Low, \textit{PR:} None,
\textit{UI:} None'' from Table \ref{tab: exploitabilityDockerHub}. 
\begin{table}[]
\caption{Distribution of base-image vulnerabilities based on exploitability metrics}
\label{tab: exploitabilityDockerHub}
\begin{threeparttable}
\scalebox{0.75}{
\begin{tabular}{|l|lll|lll|lll||lll|}
\hline
\multirow{2}{*}{UI} &
  AV: Net &
  \multicolumn{2}{c|}{AC} &
  AV: Ad &
  \multicolumn{2}{c|}{AC} &
  AV: L &
  \multicolumn{2}{c||}{AC} &
  AV: P &
  \multicolumn{2}{c|}{AC} \\
                   & PR   & L & H & PR   & L & H & PR   & L & H & PR   & L & H \\
                   \hline
\multirow{3}{*}{N} & N & 528 & 105  & N & 12  & 2    & N & 10  & 4    & N & 32  & 4    \\
                   & L  & 48  & 15   & L  & 3   & 1    & L  & 308 & 76   & L  & 26  & 1    \\
                   & H & 31  & 5    & H & 0   & 1    & H & 35  & 12   & H & 0   & 0    \\
                   \hline
\multirow{3}{*}{R} & N & 394 & 4    & N & 0   & 0    & N & 308 & 6    & N & 0   & 0    \\
                   & L  & 1   & 0    & L  & 1   & 0    & L  & 2   & 3    & L  & 0   & 0    \\
                   & H & 0   & 0    & H & 0   & 0    & H & 4   & 0    & H & 0   & 0   \\
\hline
\end{tabular}
}

\end{threeparttable}
\end{table}
\subsubsection{\textbf{PoC Exploit Collection}} For vulnerability management in container life-cycle, it is important to know whether an exploit targeting a given vulnerability is publicly available, since this would inform developers to unfold and prioritize their vulnerability mitigation policy. For instance, PoC exploit vulnerabilities can be used to configure the security policies for container orchestrator, such as developing `blacklist' vulnerabilities in Kubernetes. We leveraged  exploit data source, EDB \cite{exploitdb} since it is the largest publicly available exploit data source  \cite{mohallel2016experimenting,lin2018measurement}. We mapped each of our filtered vulnerabilities with EDB, and collected 74 exploits. We also consulted Metasploit \cite{metasp} and VulHub \cite{vulhub} to collect the vulnerable programs.       
\subsubsection{\textbf{Selection of Containerized Software}} To answer RQ3, we collected containerized software from GH by using the tool developed by Henkel \textit{et al.} \cite{henkel2020dataset} based on mining Dockerfiles of GH projects. We focused on mining Dockerfiles which were using only official images to host the applications, since earlier studies \cite{shu2017study, zerouali2019relation,  wist2020vulnerability, socchi2019deep} showed that official images were more secured than community type images. Our data set composed of 77,708 GitHub projects which were using official images. While gathering image names and corresponding tags, we observed 13,129 Dockerfiles providing tags during build time of containers. As it is not possible to specify which tags were provided during build time, we discarded 13,129 Dockerfiles from our consideration, which left 64,579 projects. Then we leveraged ANCHORE to perform vulnerability scanning of 4,383 base-images being used in 64,579 projects to identify the presence of HE and HI vulnerabilities. Table \ref{tab: githubbb} presents statistics over our studied GH projects.


\begin{table}[]

\caption{Popularity \& maintainability of studied projects (GH)}
\label{tab: githubbb}
\centering
\scalebox{0.8}{
\begin{tabular}{|c|c|c|c|c|}
\hline
Attributes         & Mean    & St. Dev & Min & Max   \\ \hline
Age (days)         & 2328.78 & 715.1   & 625 & 3696  \\ \hline
Last update (days) & 24.1    & 45.45   & 0   & 202   \\ \hline
Star               & 1011.8  & 2195.71 & 11  & 10634 \\ \hline
Watch              & 55.18   & 115.86  & 1   & 562   \\ \hline
Contributors       & 36.43   & 94.08   & 1   & 606   \\ \hline
\end{tabular}
}
\end{table}

\subsubsection{\textbf{Quantitative Analyses}} We statistically verified our observation in RQs by carrying out a) Chi-Square (CS) test \cite{satorra2001scaled} to verify the distributions of categorical variables differ from each other, b) non-parametric Mann-Whitney U (MWU) test \cite{mcknight2010mann} to compare the difference between distributions, c) Spearman (SP) rank correlation ($\rho$) \cite{spearman} to discover the strength between two sets of data, d) Mann-Kendall (MK) test \cite{mankendal} for analysing trends of data. The selected test had been considered since those are not affected by distribution of data and length of time-series data \cite{haque2020challenges}. Moreover, we considered 95\% confidence with $\alpha$ (significance level) being 0.05, which is a statistical significance level \cite{mcknight2010mann}. \textit{p-value} less than $\alpha$ indicates rejection of null hypothesis, which ensures that the two samples have different distributions at significance level of 0.05. Besides, we computed Cliff's Delta $\delta$ \cite{macbeth2011cliff}, a non-parametric measure for quantifying the difference of two observed samples, where the magnitude is assessed with the thresholds $\mid\delta\mid<$ 0.147 is `negligible', $\mid\delta\mid<$ 0.33 is `small', 
$\mid\delta\mid<$ 0.474 is `medium', and otherwise `large'. In addition, 0.1 $\leq\rho\leq$ 0.29 represent a weak, 0.3 $\leq\rho\leq$ 0.49 represent a medium, and $\rho\geq$ 0.5 represent a strong correlation. 
The data for reproducing our empirical results are provided in \cite{github}. 

\section{Results and Discussion}
\vspace{- 5 pt}
\label{sec: results}
This section presents results of our empirical investigation. 
\subsection{RQ Analysis}
\label{sec: rq1}
\subsubsection{\textbf{Exploitability of base-image vulnerabilities}} 



\begin{table}[]
\caption{Characteristics of base-image vulnerabilities (\%) in terms of attacker capabilities, HE, and HI metrics}
\label{tab: attackerCap}
\scalebox{0.75}{
\begin{tabular}{|c|c|c|c|c|c|c|c|c|}
\hline
Repository & alpine & bash & busybox & centos & debian & ibmjava & oraclelinux & ubuntu \\ \hline
External     & 84.9   & 85.1 & 87.0    & 70.2   & 57.8   & 54.2    & 91.3        & 64.8   \\ \hline
Internal     & 15.1   & 14.9 & 13.0    & 27.6   & 36.6   & 41.3    & 8.7         & 31.8   \\ \hline \hline
HE	& 59.7	& 60.4	& 58.7	& 45.1	& 42.7	& 32.3	& 33.8	& 43.6 \\ \hline
HI	& 20.1	& 26.1	& 26.7	& 39.4	& 46.6	& 18.3	& 17.3	& 40.7 \\ \hline
\end{tabular}
}
\end{table}
Table \ref{tab: attackerCap} shows the exploitability characteristics in terms of attacker capabilities of the base-image vulnerabilities across base-image repositories. Table \ref{tab: attackerCap} denotes that \textit{oraclelinux} and \textit{ibmjava} vulnerabilities are observed more in external and internal capabilities, respectively. We verified our observation with MWU test which rejected the null hypothesis of similar distribution between each pair of observed samples ($|\delta|$ $\leq$0.22, small effect size). CVE-2018-12404 (or ELSA-2019-2237) is the most frequent vulnerability in \textit{oraclelinux}, which is prevalent in all of the \textit{oraclelinux} images (tag $\geq$ 7.x). In particular, 66.7\% of \textit{oraclelinux} images and 12.2\% of all vulnerable packages are affected with this vulnerability. An attacker can launch a side channel attack during handshakes using RSA encryption to exploit the vulnerability. 
Besides, CVE-2018-7738 is found to be the most frequent vulnerability in \textit{ibmjava}, affecting 80\% of \textit{ibmjava} images and 11.6\% of vulnerable packages. In this vulnerability, unmount or bash-completion allows local users to gain privileges by embedding shell commands in a mountpoint name, which is mishandled during a unmount command (within Bash script).   
Using CS test (\textit{p}$<$0.001), we verified that base-image repositories with minimal packages, such as \textit{alpine}, \textit{bash}, and \textit{busybox} and large Operating System (OS) base-images, such as \textit{debian}, \textit{ubuntu}, \textit{centos} shown distinctive properties, e.g., vulnerabilities in minimal base-image repositories are more exposed to external attackers, whereas  vulnerabilities in  large operating system base-image repositories are more exposed to internal attackers. 

\noindent\fbox{%
    \parbox{0.98\linewidth}{%
        \textbf{Finding 1} Base-image vulnerabilities in minimal repositories are observed more in external attacker capabilities. 
    }%
}

\begin{table}[]
\centering 
\caption{Exploitability of base-image vulnerabilities}
\label{tab: attackerCap2}
\scalebox{0.75}{
\begin{tabular}{llccc|cc|cc}

\hline
 &
   &
  \multicolumn{3}{c|}{PR (\%)} &
  \multicolumn{2}{c|}{AC (\%)} &
  \multicolumn{2}{c}{UI (\%)} \\
 &
  Repository &
  None &
  Low &
  High &
  Low &
  High &
  None &
  Required
   \\ \hline
\multirow{8}{*}{\begin{turn}{90} \specialcell{External Attack} \end{turn}} &
  alpine &
  98.1 &
  1.9 &
  0 &
  74.3 &
  25.7 &
  98.1 &
  1.9  \\
 &
  bash &
  97.5 &
  2.5 &
  0 &
  100 &
  0 &
  50 &
  50  \\
 &
  busybox &
  87.5 &
  12.5 &
  0 &
  80 &
  20 &
  100 &
  0  \\
 &
  centos &
  94.9 &
  4.1 &
  0.9 &
  84.1 &
  15.9 &
  81.5 &
  18.5  \\
 &
  debian &
  99.5 &
  0.5 &
  0 &
  85.4 &
  14.6 &
  83.7 &
  16.3  \\
 &
  ibmjava &
  100 &
  0 &
  0 &
  76.8 &
  23.2 &
  78.5 &
  21.5 \\
 &
  oraclelinux &
  100 &
  0 &
  0 &
  90.5 &
  9.5 &
  61.9 &
  38.1 \\
 &
  ubuntu &
  98.7 &
  1.3 &
  0 &
  84.9 &
  15.1 &
  78.4 &
  21.6  \\ \hline
\multirow{8}{*}{\begin{turn}{90} \specialcell{Internal Attack} \end{turn}} &
  alpine &
  \multicolumn{1}{c}{11.1} &
  \multicolumn{1}{c}{88.9} &
  \multicolumn{1}{c|}{0} &
  \multicolumn{1}{c}{22.2} &
  \multicolumn{1}{c|}{77.8} &
  \multicolumn{1}{c}{88.9} &
  \multicolumn{1}{c}{11.1} \\
 &
  bash &
  \multicolumn{1}{c}{100} &
  \multicolumn{1}{c}{0} &
  \multicolumn{1}{c|}{0} &
  \multicolumn{1}{c}{100} &
  \multicolumn{1}{c|}{0} &
  \multicolumn{1}{c}{0} &
  \multicolumn{1}{c}{100} \\
 &
  busybox &
  \multicolumn{1}{c}{91.7} &
  \multicolumn{1}{c}{8.3} &
  \multicolumn{1}{c|}{0} &
  \multicolumn{1}{c}{100} &
  \multicolumn{1}{c|}{0} &
  \multicolumn{1}{c}{8.3} &
  \multicolumn{1}{c}{91.7} \\
 &
  centos &
  \multicolumn{1}{c}{35.8} &
  \multicolumn{1}{c}{62.5} &
  \multicolumn{1}{c|}{1.7} &
  \multicolumn{1}{c}{71.8} &
  \multicolumn{1}{c|}{28.2} &
  \multicolumn{1}{c}{59.8} &
  \multicolumn{1}{c}{40.2} \\
 &
  debian &
  \multicolumn{1}{c}{12.5} &
  \multicolumn{1}{c}{77.4} &
  \multicolumn{1}{c|}{10.1} &
  \multicolumn{1}{c}{71.9} &
  \multicolumn{1}{c|}{28.1} &
  \multicolumn{1}{c}{82.9} &
  \multicolumn{1}{c}{17.1} \\
 &
  ibmjava &
  \multicolumn{1}{c}{8.6} &
  \multicolumn{1}{c}{87.7} &
  \multicolumn{1}{c|}{3.7} &
  \multicolumn{1}{c}{73.7} &
  \multicolumn{1}{c|}{26.3} &
  \multicolumn{1}{c}{85.9} &
  \multicolumn{1}{c}{14.1} \\
 &
  oraclelinux &
  \multicolumn{1}{c}{50} &
  \multicolumn{1}{c}{50} &
  \multicolumn{1}{c|}{0} &
  \multicolumn{1}{c}{50} &
  \multicolumn{1}{c|}{50} &
  \multicolumn{1}{c}{50} &
  \multicolumn{1}{c}{50} \\
 &
  ubuntu &
  \multicolumn{1}{c}{17.5} &
  \multicolumn{1}{c}{76.5} &
  \multicolumn{1}{c|}{6} &
  \multicolumn{1}{c}{73.8} &
  \multicolumn{1}{c|}{26.2} &
  \multicolumn{1}{c}{80.9} &
  \multicolumn{1}{c}{19.1} \\ \hline
\end{tabular}
}
\end{table}


Table \ref{tab: attackerCap2} presents a deep inspection in the exploitability characteristics. From Table \ref{tab: attackerCap2}, it is evident that base-image vulnerabilities in \textit{busybox} is different for \textit{low} distribution in terms of PR, \textit{bash} and \textit{oraclelinux} in terms of AC (MWU test, $|\delta|\geq$0.22, small effect size) for external attacker capabilities. For UI, we did not find different distribution for \textit{alpine-busybox} and \textit{bash-oraclelinux} distribution. We also observed similar distribution for \textit{cetos}, \textit{debian}, and \textit{ubuntu} in external attacker capabilities. Besides, \textit{bash} and \textit{busybox} vulnerabilities were shown different distribution than other distributions (MWU test, $|\delta|\geq$0.22, small effect size) for internal attacker capabilities for UI metrics. Interestingly, we found that although \textit{bash} vulnerabilities did not require any privileges and minimal effort to exploit, however, they need user interaction in order to exploit in terms of internal attacker capabilities. Some common vulnerabilities in \textit{bash} images are CVE-2017-16879 (out-of-bound write), CVE-2018-10754 (improper restriction), and CVE-2017-10684 (memory overflow). These vulnerabilities can be exploited by an attacker through enticing the user to process untrusted terminal interface data, leading to potential execution of arbitrary code or denial of service.  

\noindent\fbox{%
    \parbox{0.98\linewidth}{%
        \textbf{Finding 2} All the vulnerabilities in \textit{bash} images require user interaction to exploit in internal attacker capabilities. 
    }%
}
We investigated the presence of HE vulnerabilities in the studied repositories, where we found such vulnerabilities were exhibited more in minimal base-images than large OS base-images (CS test, \textit{p-value}$<$0.001). There was a median of 12 HE vulnerabilities in minimal base-images and 8 HE vulnerabilities in the large OS base-images. Among all the base-image repositories, \textit{alpine} had the highest proportion (59.7\%) of HE vulnerabilities. The most frequent HE vulnerability in \textit{alpine} base-images is CVE-2019-14697 (affected 66.7\% images, 53.3\% of vulnerable packages). The library \textit{musl libc} ($\geq$ 1.1.23) is responsible for this vulnerability and use of this library could introduce out-of-bounds writes.  

\noindent\fbox{%
    \parbox{0.98\linewidth}{%
        \textbf{Finding 3} HE vulnerabilities are observed more in minimal base-images. 
    }%
}

We observed an increasing trend of HE vulnerabilities in large OS base-images (MK test, \textit{p-value}$<$0.001, \textit{z-value}$<$4.02). The reason can be explained with the increasing number of packages, files, libraries, and binaries used in the large OS base-images. For instance, the median number of packages being installed in the latest five releases of large OS base-images and minimal base-images were 188 and 14, respectively. Moreover, we found static trend for rest of the base-image repositories. 
\noindent\fbox{%
    \parbox{0.98\linewidth}{%
        \textbf{Finding 4} HE vulnerabilities in large OS base-images are showing an increasing trend. 
    }%
}

We investigated the relationship between HE, maintainability and popularity, and observed no significant correlation in terms of star (SP test, $\rho$=0.52, \textit{p}=0.18), maintainer (SP test, $\rho$=0.34, \textit{p}=0.4), or contributor (SP test, $\rho$=0.5, \textit{p}=0.2). However, we obtained strong correlation for download (SP test, $\rho$=0.73, \textit{p}=0.03). The reason can be explained as the larger number of downloads (Table \ref{tab: popularity}) and HE vulnerabilities (Table \ref{tab: attackerCap}) were observed in minimal base-image repositories.

\noindent\fbox{%
    \parbox{0.98\linewidth}{%
        \textbf{Finding 5} There is a strong correlation between high exploitability and popularity (in terms of downloads). 
    }%
}



\subsubsection{\textbf{Impact of base-image vulnerabilities}}

\begin{table}[]
\centering
\caption{Impact of base-image vulnerabilities}
\label{tab:immm}
\scalebox{0.75}{
\begin{tabular}{llccc|ccc|ccc}
\hline
 &
   &
  \multicolumn{3}{c|}{Confidentiality (\%)} &
  \multicolumn{3}{c|}{Integrity (\%)} &
  \multicolumn{3}{c}{Availability (\%)} \\
 &
  Repository &
  None &
  Low &
  High &
  None &
  Low &
  High &
  None &
  Low &
  High \\ \hline
\multirow{8}{*}{\begin{turn}{90} \specialcell{External Attack} \end{turn}} &
  alpine &
  22.8 &
  35.6 &
  41.6 &
  70.3 &
  0 &
  29.7 &
  55.4 &
  0 &
  44.6 \\
 &
  bash &
  72.5 &
  0 &
  27.5 &
  72.5 &
  0 &
  27.5 &
  0 &
  0 &
  100 \\
 &
  busybox &
  11.3 &
  0 &
  88.7 &
  46.3 &
  0 &
  53.7 &
  35 &
  0 &
  65 \\
 &
  centos &
  46.4 &
  7.9 &
  45.6 &
  58.5 &
  6.3 &
  35.2 &
  23.8 &
  1.6 &
  74.5 \\
 &
  debian &
  45.9 &
  10.4 &
  43.7 &
  56.7 &
  2.9 &
  40.4 &
  17.5 &
  0.7 &
  81.9 \\
 &
  ibmjava &
  46.1 &
  15.9 &
  38 &
  61.6 &
  3.46 &
  34.9 &
  15.9 &
  10.3 &
  73.7 \\
 &
  oraclelinux &
  38.1 &
  38.1 &
  23.8 &
  42.8 &
  28.6 &
  28.6 &
  52.4 &
  0 &
  47.6 \\
 &
  ubuntu &
  54.9 &
  7 &
  38.1 &
  56.4 &
  3.9 &
  39.7 &
  14.2 &
  3.2 &
  82.6 \\ \hline
\multirow{8}{*}{\begin{turn}{90} \specialcell{Internal Attack} \end{turn}} &
  alpine &
  \multicolumn{1}{l}{22.2} &
  \multicolumn{1}{l}{0} &
  \multicolumn{1}{l|}{77.8} &
  \multicolumn{1}{l}{88.9} &
  \multicolumn{1}{l}{0} &
  \multicolumn{1}{l|}{11.1} &
  \multicolumn{1}{l}{88.9} &
  \multicolumn{1}{l}{0} &
  \multicolumn{1}{l}{11.1} \\
 &
  bash &
  \multicolumn{1}{l}{14.3} &
  \multicolumn{1}{l}{0} &
  \multicolumn{1}{l|}{85.7} &
  \multicolumn{1}{l}{14.3} &
  \multicolumn{1}{l}{0} &
  \multicolumn{1}{l|}{85.7} &
  \multicolumn{1}{l}{0} &
  \multicolumn{1}{l}{0} &
  \multicolumn{1}{l}{100} \\
 &
  busybox &
  \multicolumn{1}{l}{100} &
  \multicolumn{1}{l}{0} &
  \multicolumn{1}{l|}{0} &
  \multicolumn{1}{l}{8.3} &
  \multicolumn{1}{l}{0} &
  \multicolumn{1}{l|}{91.7} &
  \multicolumn{1}{l}{8.3} &
  \multicolumn{1}{l}{0} &
  \multicolumn{1}{l}{91.7} \\
 &
  centos &
  \multicolumn{1}{l}{41.3} &
  \multicolumn{1}{l}{7.7} &
  \multicolumn{1}{l|}{50.1} &
  \multicolumn{1}{l}{62.2} &
  \multicolumn{1}{l}{2.7} &
  \multicolumn{1}{l|}{35.1} &
  \multicolumn{1}{l}{25.9} &
  \multicolumn{1}{l}{4.76} &
  \multicolumn{1}{l}{69.3} \\
 &
  debian &
  \multicolumn{1}{l}{34.9} &
  \multicolumn{1}{l}{1.3} &
  \multicolumn{1}{l|}{63.8} &
  \multicolumn{1}{l}{27.6} &
  \multicolumn{1}{l}{0.4} &
  \multicolumn{1}{l|}{72.1} &
  \multicolumn{1}{l}{11.2} &
  \multicolumn{1}{l}{3.4} &
  \multicolumn{1}{l}{85.1} \\
 &
  ibmjava &
  \multicolumn{1}{l}{19.5} &
  \multicolumn{1}{l}{19.1} &
  \multicolumn{1}{l|}{61.4} &
  \multicolumn{1}{l}{21.8} &
  \multicolumn{1}{l}{16.4} &
  \multicolumn{1}{l|}{61.8} &
  \multicolumn{1}{l}{20.4} &
  \multicolumn{1}{l}{16.4} &
  \multicolumn{1}{l}{63.2} \\
 &
  oraclelinux &
  \multicolumn{1}{l}{0} &
  \multicolumn{1}{l}{50} &
  \multicolumn{1}{l|}{50} &
  \multicolumn{1}{l}{50} &
  \multicolumn{1}{l}{0} &
  \multicolumn{1}{l|}{50} &
  \multicolumn{1}{l}{0} &
  \multicolumn{1}{l}{50} &
  \multicolumn{1}{l}{50} \\
 &
  ubuntu &
  \multicolumn{1}{l}{24.6} &
  \multicolumn{1}{l}{10.5} &
  \multicolumn{1}{l|}{64.9} &
  \multicolumn{1}{l}{25.1} &
  \multicolumn{1}{l}{6.4} &
  \multicolumn{1}{l|}{68.5} &
  \multicolumn{1}{l}{17.5} &
  \multicolumn{1}{l}{6.8} &
  \multicolumn{1}{l}{75.7} \\ \hline
\end{tabular}
}
\end{table}

Table VII shows the effect on the \textit{CIA} triad of the studied vulnerabilities. We observed a similar distribution of vulnerabilities of `High' value in \textit{Confidentiality} except \textit{busybox} for external attack and `None' value for internal attack (MWU test, $|\delta|\geq$0.22, small effect size). 
Interestingly, \textit{busybox} vulnerabilities could not result in any loss of information if the vulnerability is exploited from internal attacker capabilities. For \textit{Integrity} metrics, external attacker capabilities exhibit similar distribution for `High' and `None' values. On the other hand, \textit{alpine} showed less impacted for internal attackers for data \textit{integrity} in case of successful exploitation. However, \textit{bash} images are the most affected than other base-image vulnerabilities, since they can cause complete inaccessibility of the hosted applications for both attacker capabilities if the exploitation is successful.

\noindent\fbox{%
    \parbox{0.98\linewidth}{%
        \textbf{Finding 6} Exploitation of \textit{bash} vulnerabilities can result in complete unavailability of the impacted container. 
    }%
}
Our investigation for the presence of HI vulnerabilities in the studied repositories revealed that large OS base-images had higher appearance than other base-images (CS test, \textit{p-value}$<$0.001). There was a median of 4 HI vulnerabilities in minimal base-images and 10 HI vulnerabilities in large OS base-images. Among all repositories, \textit{debian} had the highest proportion (46.9\%) of HI vulnerabilities. The most frequent HI vulnerability in \textit{debian} base-images is CVE-2015-5224 (affected 82.4\% images, 40.3\% of vulnerable packages). The library \textit{login-utils in util-linux} ($\geq$ 2.29) is responsible for this vulnerability and allows remote attackers to cause file name collision resulting in complete loss of user information.
\noindent\fbox{%
    \parbox{0.98\linewidth}{%
        \textbf{Finding 7} HI vulnerabilities are observed more in large OS base-images.  
    }%
}

We discovered an increasing trend of HI vulnerabilities in \textit{ubuntu} (MK test, \textit{p-value}$=$0.001, \textit{z-value}$=$6.73), \textit{ibmjava} (MK test, \textit{p-value}$=$0.004, \textit{z-value}$=$2.98), and static trend for rest of repositories. The most common HI vulnerability in \textit{ubuntu} images is CVE-2016-2779 (affected 61.4\% images, 13.6\% packages). Several packages are affected with this vulnerability, such as \textit{util-linux}, \textit{libuuid}, \textit{libblkid}, and \textit{libmount}. The exploitation can cause escalation of privileges for impacted container.  
\noindent\fbox{%
    \parbox{0.98\linewidth}{%
        \textbf{Finding 8} HI vulnerabilities in \textit{ubuntu} and \textit{ibmjava} are showing an increasing trend.  
    }%
}
We investigated the relationship between HI, maintainability, popularity, and observed no significant correlation in terms of star (SP test, $\rho$=0.02, \textit{p}=0.95), download (SP test, $\rho$=0.19, \textit{p}=0.65), maintainer ($\rho$=0.67, \textit{p}=0.06) or contributor (SP test, $\rho$=-0.17, \textit{p}=0.68). Similarly, no correlation is observed between HI and HE (SP test, $\rho$=-0.26, \textit{p}=0.63).  
\noindent\fbox{%
    \parbox{0.98\linewidth}{%
        \textbf{Finding 9} No correlation is found between HI and popularity, HI and maintainability or HI and HE vulnerabilities.  
    }%
}

\begin{table}[]
\centering
\caption{Top-3 base-image repositories with PoC exploitation}
\label{tab: pocex}
\scalebox{0.75}{
\begin{tabular}{|c|c|c|c|}
\hline
Repository & PoC (\%) & Base-image (\%) & Example                                      \\ \hline
ubuntu     & 7.4      & 57.3            & ubuntu:14.04 (CVE-2015-7855, EDB-ID: 40840 ) \\ \hline
debian     & 8.6      & 74.3            & debian:jessie (CVE-2018-1124, EDB-ID: 44806) \\ \hline
centos     & 5.1      & 90.3            & centos:5 (CVE-2016-2776, EDB-ID: 40453)      \\ \hline
\end{tabular}
}
\end{table}
\subsubsection{\textbf{Prevalence of vulnerable base-images}} \label{rq3} In our analysis, we found 72.8\% of base-images in DH contain at least one vulnerability with HE or HI. Moreover, we investigated the availability of fixes for the remediation of vulnerabilities and found 68.4\% of them do not have any available fixes. We calculated the exposure time \cite{liu2020understanding} of such vulnerabilities and found these vulnerabilities remain unfixed for 796 days on average, which may be adequate time-frame for attackers to exploit. Liu \textit{et al.} showed that the vulnerabilities in a common software remain unfixed for 181 days on average \cite{liu2020understanding}, which is less than the vulnerabilities in containerized context. Interestingly, we also found the presence of vulnerabilities with PoC exploit as shown in Table \ref{tab: pocex}. PoC exploit vulnerabilities in base-images can make applications hosted by the software infrastructure (built upon the base-images) highly prone to security attacks. In addition, these PoC vulnerabilities are also found to be existed in dependent official images.

\noindent\fbox{%
    \parbox{0.98\linewidth}{%
        \textbf{Finding 10} Nearly half of DH official base-images contain at least a vulnerability with PoC exploit.    
    }%
}

The reason can be explained with the large appearance of stale base-images in DH repositories. We found only a negligible amount of base-images (median 11) are being updated in DH regularly. We did not find any correlation between stale images and popularity, such as star (SP test, $\rho$=0.15, \textit{p}=0.7), downloads (SP test, $\rho$=0.45, \textit{p}=0.26), or one of the maintainability attributes, i.e.,  contributor (SP test, $\rho$=0.5, \textit{p}=0.36). However, we obtained a strong correlation between maintainer and stale images ($\rho$=-0.76, \textit{p}=0.02) which implies larger maintenance teams can better support the security of base-images on a regular basis.       

\noindent\fbox{%
    \parbox{0.98\linewidth}{%
        \textbf{Finding 11} There is a strong correlation between stale base-images and maintainer.}%
}
Moreover, we also focused on the identification of base-images with zero vulnerabilities, which is expected to have paramount importance, since applications built upon such base-images with zero vulnerabilities will inherit no vulnerabilities (from base-image). Thus, the exploitation and impact of vulnerabilities will be minimum. In this regard, we identified base-images with zero vulnerabilities and obtained 72 base-images. In addition, we also discovered 77 images derived from secure base-images which contain no vulnerabilities. We discovered
\textit{alpine} (7.4\%),
\textit{amazonlinux} (5.4\%), and
\textit{busybox} (4.7\%) constituted top-3 base-image repositories with no vulnerabilities. Besides, \textit{elixir} (5.4\%), \textit{python} (4.7\%), \textit{rabbitmq} (4.7\%), and
\textit{haproxy} (4\%) made up the top-3 non base-image repositories with zero vulnerabilities. The complete list of vulnerability-free base-images are provided in our reproduction package \cite{github}. We observed these base-images were lightweight and thus application built with such base-images could keep application size small (Section \ref{section: impact}). 
Apart from DH, we examined the composition of base-image vulnerabilities in terms of exploitation and impact, that were being used to host containerized applications in GH. We identified 91.6\% of containerized applications based on a vulnerable base-image, which contain at least one vulnerability with HE or HI, and rest of 8.4\% of containerized applications hosted in vulnerability-free base-images. This demonstrated different distribution of container applications built upon vulnerable base-images and vulnerability free base-images (MWU test, $|\delta|\geq$ 0.26, small effect size). Table \ref{tab: contapp} shows the top-3 base-images across all the studied GH projects hosting containerized applications along with their exploitability and impact.
\noindent\fbox{%
    \parbox{0.98\linewidth}{%
        \textbf{Finding 12} Majority of GH projects based on  images with HE or HI vulnerabilities.}%
}

\begin{table}[]
\caption{Top-3 images hosting container applications in GH}
\centering
\label{tab: contapp}
\scalebox{0.75}{
\begin{tabular}{|c|c|c|c|c|}
\hline
Image &
  \#application &
  \begin{tabular}[c]{@{}c@{}}HE (\%)\end{tabular} &
  \begin{tabular}[c]{@{}c@{}}HI(\%)\end{tabular} &
  \begin{tabular}[c]{@{}c@{}}HE and HI (\%)\end{tabular} \\ \hline
ubuntu:16.04  & 4,428 & 40 & 35 & 15 \\ \hline
ubuntu:14.04  & 3,680 & 46.4 & 47.8 & 28.2 \\ \hline
debian:jessie & 2,170 & 76.7 & 30 & 23.3 \\ \hline
\end{tabular}
}
\end{table}
 
We investigated the dependence of vulnerable and secure base-images in terms of GH attributes, such as popularity (star and watch count), maintainability (contributors and owner type) and project age. Interestingly, we found statistically significant difference for the GH projects using secure base-images in terms of organization owned project and large contributors (MWU test, $|\delta|\geq$ 0.26, small effect size). This could be interpreted as the organization projects or large team size have better security management than user maintained projects.   
\noindent\fbox{%
    \parbox{0.98\linewidth}{%
        \textbf{Finding 13} GH projects owned by organizations or have large contributors are more likely to use secure images.  }%
}

Besides, we observed vulnerable base-images were not upgraded or changed, i.e., once the application left the development phase, images upon which the application based on, received nominal focus. Our observation is also aligned with Cito \textit{et al.} and Lin \textit{et al.}, where they showed base-images were the least frequently changed instructions in Dockerfiles \cite{cito2017dockerfiles, lin2020large}. However, a single exploitation of base-image vulnerabilities could open the potential of security attacks and therefore upgrading to secured base-images with minimal exploitation and impact in container life-cycle can help the reduction of possibilities of security attack in containerized application.

\vspace{- 5 pt}
\section{Usefulness of Secured Base-Image}
\label{section: impact}
The purpose of this section is to discuss the usefulness of utilizing secured base-images  that we had empirically discovered by our study. We built the bulletin board application \cite{nodebulletinboard}, described in Section \ref{section: reseachProblem} using \textit{node:alpine} as a base-image. Using ANCHORE, only 4 vulnerabilities were obtained for the final image. Since \textit{node:alpine} image had no vulnerabilities, use of vulnerability-free base-image reduced the vulnerability in the final image drastically, compared to 69 vulnerabilities reported in Section \ref{section: reseachProblem}. Consequently, vulnerability in the application remained 4 with the use of secure base-image, \textit{node:alpine}. Moreover, in line with reduction of vulnerabilities, the application size decreased to 127 MB (using \textit{node:alpine} as a base-image) from 183 MB (using \textit{node:current-slim} as a base-image) which is aligned with the containerization goal of reducing the virtualization size \cite{cito2017dockerfiles}. 

\begin{figure}[]
    \centering
    \includegraphics[height = 1.8 in]{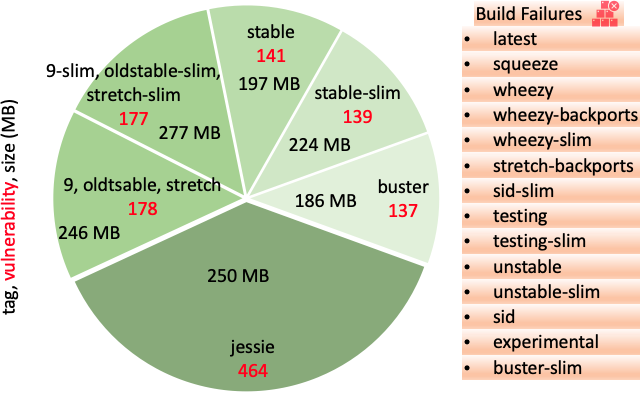}
    \caption{Vulnerability and Build Failures with  base-images.}
    \label{fig: rceee}
\end{figure}

However, it is important to note that all tags of a secure base-image may not be used to build an application. We provide an example to support our claim. For instance, consider a Dockerfile \cite{rce} which exhibits the behaviour of sending emails. Base-image \textit{debian:jessie} had been used to build the application and 464 vulnerabilities were obtained by using ANCHORE. We tried to build the same application with 26 different tags of \textit{debian}, which had fewer vulnerabilities than \textit{debian:jessie}. The vulnerability composition and build status with different base-images are shown in Fig. \ref{fig: rceee}. It is evident from Fig. \ref{fig: rceee} that the vulnerabilities had been reduced to 137 for \textit{debian:buster} from 464, indicating a larger portion of vulnerabilities; in fact 327 vulnerabilities were inherited from base-image in this application. More importantly, this inherited vulnerabilities can be eliminated with use of secure base-images. We computed the number of both HE and HI vulnerabilities in the application built from \textit{debian:jessie} and \textit{debian:buster} and found that such vulnerabilities had been reduced by 88.2\%. This demonstration showed that secured base-images can ensure two important objectives of virtualization \cite{cito2017dockerfiles}. Firstly, secured base-images propagate less vulnerabilities to applications, thus the attack surface in terms vulnerability exploitation and impact regarding \textit{CIA} can be kept minimum. Secondly, using secure base-images are aligned with the goal of containerization to reduce the virtualization footprint. However, we can observe all the tags for secured \textit{debian} base-images can not build the application, which is shown as `Build Failures' in Fig. \ref{fig: rceee}. Build Failures represent that the application can not be built using those tags due to dependency issues, i.e., unavailability of some packages. 
\vspace{- 8 pt}  
\section{Implications}
\label{sec: discussion}
In this Section, we provide some implications of our work for researchers, relevant practitioners and Docker developers.
    
\subsubsection{\textbf{For Practitioners}}
\label{section: practitionersDiscussion}
Our vulnerability data set along with its exploitation and impact metrics can help practitioners to select secure base-images for building containerized applications. Moreover, practitioners can check the vulnerabilities in different tags and avoid using base-images with HE vulnerabilities since such vulnerabilities can initiate other attacks (\textbf{RQ1}). Practitioners need to take specific configuration for their base-images. For instance, applications hosting upon minimal images should harden their software-defined networks (\textbf{RQ1}). On the other hand, large OS images inherited applications should pay more attention in local access, authentication, and privileges of users.  In addition, our research calls for effort on the release of patches for non-fixed vulnerabilities (\textbf{RQ3}). Practitioners can use lightweight base-images to keep the application size small; consequently vulnerabilities with HE and HI could be minimized (\textbf{RQ3}). Moreover, our findings suggest that if a project has a larger community in terms of developers or contributors, then the likelihood of using vulnerable images are reduced, since more developers can focus on secured development and maintenance of containers. 
\subsubsection{\textbf{For Container Developers}} Docker base-image developers can consider reducing the transitive dependencies, which can minimize vulnerability propagation. Besides, they need to prioritize their remediation process by considering the exploitability and impact metrics (\textbf{RQ1, RQ2}). Moreover, they can consider generating warning messages while users attempt to download stale images from repositories so that it can alert the image users for subsequent security risk (\textbf{RQ3}). Docker developers should also avoid base-images with PoC exploit vulnerabilities for instantiating other official images (\textbf{RQ3}). 
\subsubsection{\textbf{For Security Tool Builders}} Security tool builders need to include exploitability metrics so that the security report can be contextualised in container technologies (\textbf{RQ1}). Moreover, they can also add exposure time to the unfixed vulnerabilities which can help users to select secured base-images (\textbf{RQ2}). Tool builders may develop support tools to ensure the dependency and consistency of containerized application, such as recommending security-focused base-images based on functionality (Section \ref{section: impact}).
\subsubsection{\textbf{For Researchers}} Our data set can be used in future research to discover the trend-line of vulnerabilities across different tags of base-images. Our data set will be helpful to identify if new HE vulnerabilities are incorporated due to release of new base-images (\textbf{RQ1}). Researchers can investigate the patterns of fixed vulnerabilities in packages and leverage machine learning methods to predict fixes for long exposed HI vulnerabilities (\textbf{RQ2}). We stimulate the need of integration of security and configuration tools to ensure configuration and compliance of containers in their life-cycle (\textbf{RQ3}).
\vspace{- 5 pt}
\section{Threats to validity}
\label{sec: threats}
\subsubsection{\textbf{Construct Validity}} 
To provide exploitability characteristics of base-images, we need to create a base-image specific data set. Images hosted in base-image repositories of DH do not mention if an image has been used as a base-image to build another images. Therefore, creating a base-image specific data set becomes a challenge as well as a threat to our study. To mitigate the threat, we leveraged all the official images in DH. Then we crawled corresponding Dockerfiles and considered $FROM$ statement to collect exact image names and tags used in Dockerfiles. We iteratively mined image names unless it fell in \textit{scratch}. This allowed us to create a data set of base-images along with their tags, used to build official images, which are further extended by container developers.    
\subsubsection{\textbf{Internal Validity}}
Our results may vary if different static tools and CVSS version (e.g., version 2) are used. However, usage of single static tool is common in the existing related literature \cite{shu2017study, zerouali2019relation, wist2020vulnerability, socchi2019deep, liu2020understanding}. Moreover, we used CVSS version 3, which is designed to overcome the shortcoming of CVSS version 2 \cite{cvss}. Our threat model did not consider the exploitation of base-image vulnerabilities which require hardware resource manipulation or physical access. However, our threat model was conceptualised based on prior research for container security \cite{sultan2019container, modi2017virtualization,  lin2018measurement}. We used EDB \cite{exploitdb} to collect PoC exploit base-image vulnerabilities and might miss PoC exploit vulnerabilities if there was no entry on EDB. EDB is the largest publicly available exploit data sources and also used by prior research \cite{mohallel2016experimenting, lin2018measurement, tunde2019study}. Furthermore, we also collected vulnerable programs from Metasploit \cite{metasp} and VulHub \cite{vulhub} to understand the exploitability and impact of base-image vulnerabilities. 
\subsubsection{\textbf{External Validity}}
While crawling from NVD, 286 out of 2,269 base-image vulnerabilities were discarded due to their unavailable exploitability and impact information in CVSS version 3. We released the data set \cite{github} for replication.


\vspace{- 5 pt}
\section{Related Work}
\label{sec: related}

Shu \textit{et al.} \cite{shu2017study} performed the first study on common vulnerabilities in images hosted in DH. They proposed a framework named Docker Image Vulnerability Analysis (DIVA) for collecting and analysing images. Wist \textit{et al.} \cite{wist2020vulnerability} also performed a similar study recently on 2,500 DH images. Moreover, Kwon \textit{et al.} \cite{kwon2020divds} proposed a Docker Image Vulnerability Diagnostic System (DIVDS) to diagnoses images when uploading or downloading the images from DH. Zerouali \textit{et al.} focused on vulnerabilities and outdated packages in \textit{debian}-based images \cite{zerouali2019relation, zerouali2021multi}, programming language images \cite{zerouali2021impact,zerouali2021usage}  from DH. Mohallel \textit{et al.} \cite{mohallel2016experimenting} showed the presence of vulnerability in images can increase attack surface for Docker than traditional virtual machines. Ibrahim \textit{et al.} showed the lack of guidance for developers to select suitable images to build applications \cite{ibrahim2020too}. Tak \textit{et al.} performed study on DH images and showed that a majority of the images contained vulnerable packages and violated security compliance rules \cite{tak2018security}. Socchi \textit{et al.} investigated the findings revealed by Shu \textit{et al.} \cite{shu2017study} and showed prevalence of vulnerabilities in all types of repositories, such as official, community, verified and certified \cite{socchi2019deep}. Besides, Liu \textit{et al.} identified user unawareness on using images from DH and proposed a framework to identify malicious images \cite{liu2020understanding}. 

However, our work is significantly different to the previous studies in several ways. Firstly, our research questions and goals of the study are different in terms of identification of exploitability of known vulnerabilities, their assessment, prioritization, remediation and  of such exploitable vulnerabilities, providing a comprehensive evidence-based knowledge on vulnerability management for base-images. We also investigated the practice of developing containerized software in terms of analyzing the relationship of vulnerable/secure base-images with different project attributes (e.g., project age, popularity, contributor size, etc) from GH. 

Secondly, the data we analyzed is significantly different from the previous studies due to our base-image extraction process. To collect base-images, we leveraged the Dockerfile from all official repositories of DH and iteratively mined images (mentioned in Dockerfile) until it fell in \textit{scratch}. In contrast, the most relevant study \cite{socchi2019deep} used only the most recent updated image (i.e., \textit{latest}-tagged images) from the base-image repositories, which may missed a huge portion of actually deployed base-images. Our assumption is verified as we found 99.1\% of official image's Dockerfiles did not use \textit{latest}-tagged images, indicating that such \textit{latest}-tagged base-images are not in the real-world practice of building official images in DH\footnote{Oracle Linux officially removed \textit{latest}-tagged images from their repository, \url{https://hub.docker.com/_/oraclelinux?tab=description}}. Moreover, our analyzed images are different (minimally overlapped, less than 1\%) with the most recent studies on image vulnerabilities \cite{zerouali2019relation, wist2020vulnerability, ibrahim2020too}, which mean the findings and observations we contributed are substantially different than the earlier studies. 

Thirdly, the vulnerability report in our study was computed using ANCHORE \cite{anchoreTool}, which showed better precision and recall than CLAIR \cite{berkovich2020ubcis}. Therefore, we suspect the vulnerabilities reported by the studies \cite{shu2017study, socchi2019deep} using CLAIR \cite{clair} may miss some vulnerabilities. We verified our observation by scanning our base-image data set with CLAIR, and found that it could not detect 28.8\% of the existing vulnerabilities, which indicates our data set contains broader set of vulnerabilities. Besides, our data set is generalized, and not specific to any particular distribution, such as \textit{debian} \cite{zerouali2019relation, ibrahim2020too}.

In summary, our work has uniquely performed in-depth vulnerability analysis of `base-images', which are used to create containerized software. This analysis is indispensable for a precise understanding of security issues in container-based applications since base-image is a key component in multi-tenant environments. Our results also provide an empirically discovered secure `base-image' library which can be used to build secure containerized applications. Moreover, we have also demonstrated how propagation of vulnerabilities from base-images can be minimized by using the identified secured base-images list.  
\vspace{- 5 pt}
\section{Conclusion and future work}

\label{sec: conclussion}
With the rapidly increasing trend of deploying applications in Docker containers, understanding the exploitation, impact, remediation and prevalence of base-image vulnerabilities in the practice has eminent importance since base-images provide all the core functionalities to build and run containerized software. A single exploitation of base-image vulnerabilities could open the floodgates for numerous security attacks in thousands of containerized software, leaving the ecosystem with countless security hazards. In this aspect, we have created a vulnerability data set with its PoC exploitation for actively maintained base-images of DH. This data set enabled us to perform an empirical study related to vulnerability exploitation, impact, availability of patches, as well as its real-world prevalence in terms of characterizing usage of DH images to host applications in GH. Our findings are expected to guide and motivate developers for secured configuration of containers in terms of using secured base-images. We provided a list of  base-images containing PoC exploitation vulnerabilities as well as base-images with zero vulnerabilities. We demonstrated the usage of secure base-images, which can reduce the attack surface and serve the purpose of optimal sizes for containerized software. We also verified our observations by means of statistical methods, such as Man Whitney U test. We plan to build a recommender system to suggest secure base-images based on the functionality and dependency of an application.

\section*{Acknowledgment}
The work has been supported by the Cyber Security Research Centre Limited whose activities are partially funded by the Australian Government’s Cooperative Research Centres Programme.
This work has also been supported with super-computing resources provided by the Phoenix HPC service at The University of Adelaide. 





\bibliographystyle{IEEEtran}
\bibliography{sample-base}
%



\end{document}